\documentclass[runningheads]{llncs}

\usepackage[T1]{fontenc}
\usepackage[utf8]{inputenc}
\usepackage{graphicx}
\usepackage{url}
\usepackage{booktabs}
\usepackage{tikz}
\usepackage{siunitx}
\usepackage{rotating}
\usepackage{microtype}
\usepackage{hyperref}
\usepackage{tabularx}

\DeclareSIUnit\dBm{dBm}
\DeclareSIUnit\byte{B}
\DeclareSIUnit\hertz{Hz}
\DeclareSIUnit\wattpeak{W_p}

\begin{document}

\title{A Meshtastic-based LoRa Mesh System for Smart Campus Applications: From Solar-Powered Sensing to Containerized Data Management}

\titlerunning{LoRa Mesh System for Smart Campus}

\author{Rafael Garzon Andosilla \inst{1} \and Jose Rugeles\inst{1}}

\institute{Universidad Militar Nueva Granada, Bogot\'{a}, Colombia\\
\email{\{est.rafael.agarzon, jose.rugeles\}@unimilitar.edu.co}}

\maketitle

\begin{abstract} 
This work presents the design, implementation, and evaluation of a LoRa-based mesh network using the Meshtastic protocol for Smart Campus applications at Universidad Militar Nueva Granada (UMNG). The system integrates heterogeneous hardware nodes — including a solar-powered ecological sensing node built around a Raspberry Pi Pico and a Semtech SX1262 transceiver, and mobile trackers based on the Seeed SenseCAP T1000-E — managed through a containerized edge gateway running on a Raspberry Pi 4. A Docker Compose microservices stack handles data ingestion via Node-RED, time-series storage in InfluxDB, and real-time visualization through Grafana dashboards. The architecture's performance was evaluated under realistic propagation scenarios at the UMNG Cajicá campus, characterizing link quality using Received Signal Strength Indicator (RSSI) and Signal-to-Noise Ratio (SNR) metrics. Experimental results demonstrate robust mesh connectivity across key university facilities, including an extended-range link of approximately 2.47 km linking the campus gateway to a remote station at Mirador La Cumbre (n = 62 packets received, mean RSSI = -110 dBm, mean SNR = +2.75 dB). This architecture demonstrates that open-source mesh protocols combined with containerized microservices offer an autonomous, highly reproducible infrastructure for environmental monitoring and asset tracking, supporting the transition toward data-driven "Smart Campus" ecosystems without reliance on centralized commercial LoRaWAN operators.
\end{abstract}


\section{Introduction}

The expansion of Internet of Things (IoT) infrastructure in academic 
environments has driven the need for low-cost, scalable, and 
self-organizing communication systems capable of operating across 
heterogeneous physical spaces without dependence on pre-existing 
cellular or Wi-Fi coverage. Long Range (LoRa) radio technology, 
combining sub-GHz operation, spread-spectrum modulation, and low 
power consumption, constitutes a strong candidate for such 
deployments. However, conventional LoRa implementations based on 
the LoRaWAN protocol impose a star topology that requires centralized 
network servers and device pre-registration procedures, limiting rapid 
or ad-hoc deployment in research environments~\cite{Huy2025},~\cite{Saadoon2024}.

Mesh-based alternatives have emerged as a response to these 
limitations. AODV-based routing over LoRa has demonstrated high 
reliability in mobile monitoring scenarios, with packet delivery 
ratios above 95\% in vehicular networks~\cite{Huy2025}. The Meshtastic 
open-source firmware extends this concept by implementing a fully 
decentralized \textit{managed flooding} algorithm featuring 
duplicate-ID caching and SNR-based retransmission prioritization, 
and has been validated in off-grid rescue operations and controlled 
academic settings~\cite{Chung2025}. At the physical layer, Spreading Factor 
(SF) selection governs the range--throughput--energy tradeoff: lower 
SF values maximize data rate, while SF11--12 extend coverage at the 
cost of increased Time-on-Air and energy consumption~\cite{Ghodhbane2024,MalikMatin2024}.
Recent work has demonstrated battery-less LoRa node operation via 
photovoltaic transducers, and Meshtastic-based solar nodes 
have been evaluated for asynchronous telemetry in semi-urban campus 
environments~\cite{Ramesh2025}.
Despite this progress, empirical characterization of Meshtastic 
under heterogeneous node configurations—combining fixed environmental 
sensing, mobile asset tracking, and containerized edge 
processing—remains limited.Prior studies have addressed isolated 
use cases or homogeneous node types, leaving open questions about 
mesh performance and end-to-end data pipeline integration at campus 
scale.

This work addresses that gap by presenting the design and 
experimental evaluation of a LoRa mesh network using Meshtastic, 
deployed at the UMNG Cajicá campus as a Smart Campus prototype. The 
system integrates two complementary use cases: (i) fixed 
environmental monitoring through a solar-powered sensing node 
equipped with a Davis~6450 solar radiation sensor, and (ii) mobile 
asset tracking using a Seeed SenseCAP T1000-E GNSS-enabled device. 
Telemetry is transported over the mesh to an edge gateway running a 
containerized microservices stack for ingestion, processing, storage, 
and visualization. The principal contributions of this work are: 
(a)~a detailed hardware implementation of a solar-powered LoRa node 
based on the Raspberry Pi Pico (RP2040) and Semtech SX1262; 
(b)~a reproducible Docker Compose deployment of the complete data 
pipeline; and (c)~an empirical characterization of Meshtastic mesh 
performance under semi-urban campus conditions.

\section{Proposed System Architecture}

\begin{sidewaysfigure}[p]
    \centering
    \includegraphics[width=\textwidth]{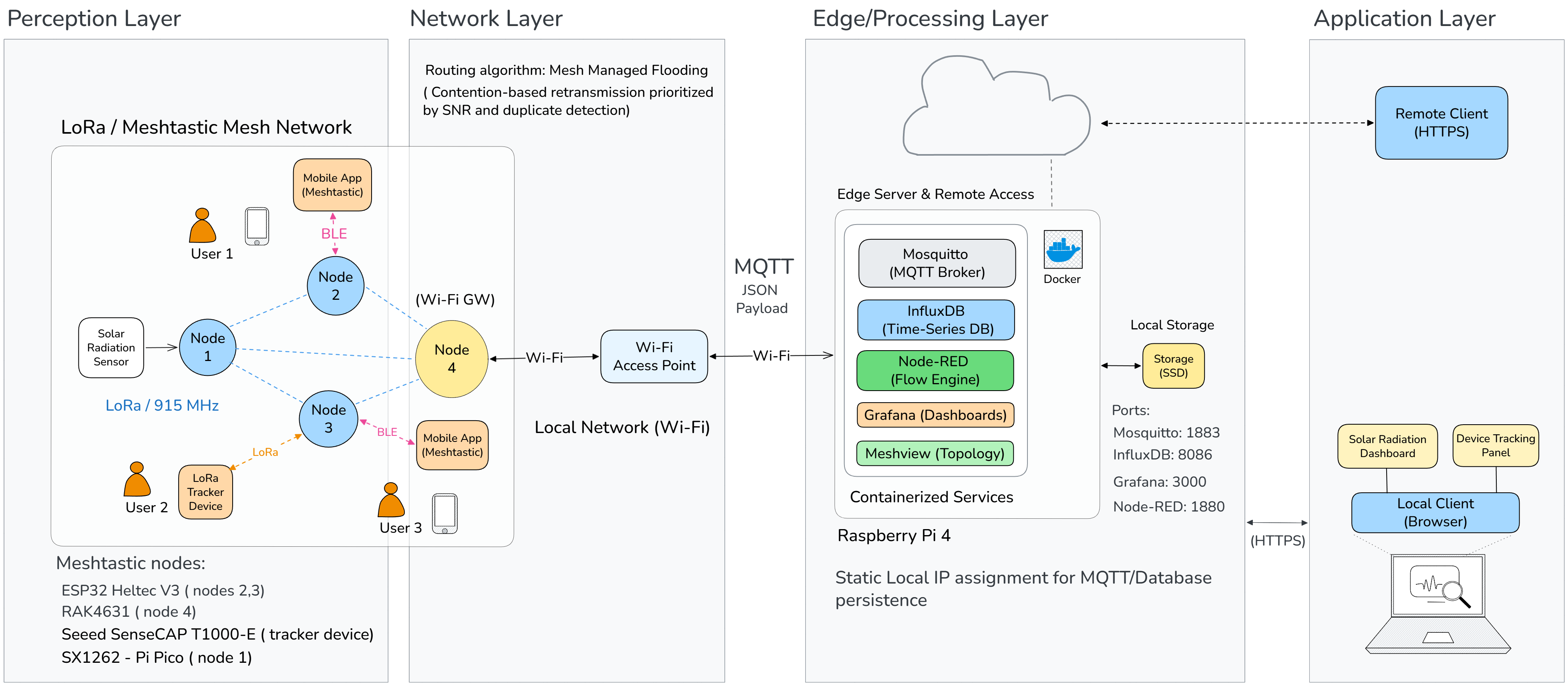}
    \caption{Full system architecture showing the functional layers: Perception, Network, Edge/Processing, and Application.}
    \label{fig:arch_full_1}
\end{sidewaysfigure}

The proposed system follows a four-layer architecture—Perception, 
Network, Edge/Processing, and Application—as illustrated in 
Fig.~\ref{fig:arch_full_1}. Each layer encapsulates a distinct 
functional domain, enabling modular deployment and independent 
scaling of sensing, routing, and data management subsystems.

\subsection{Perception Layer}

\begin{figure}[htbp]
    \centering
    \includegraphics[width=\columnwidth]{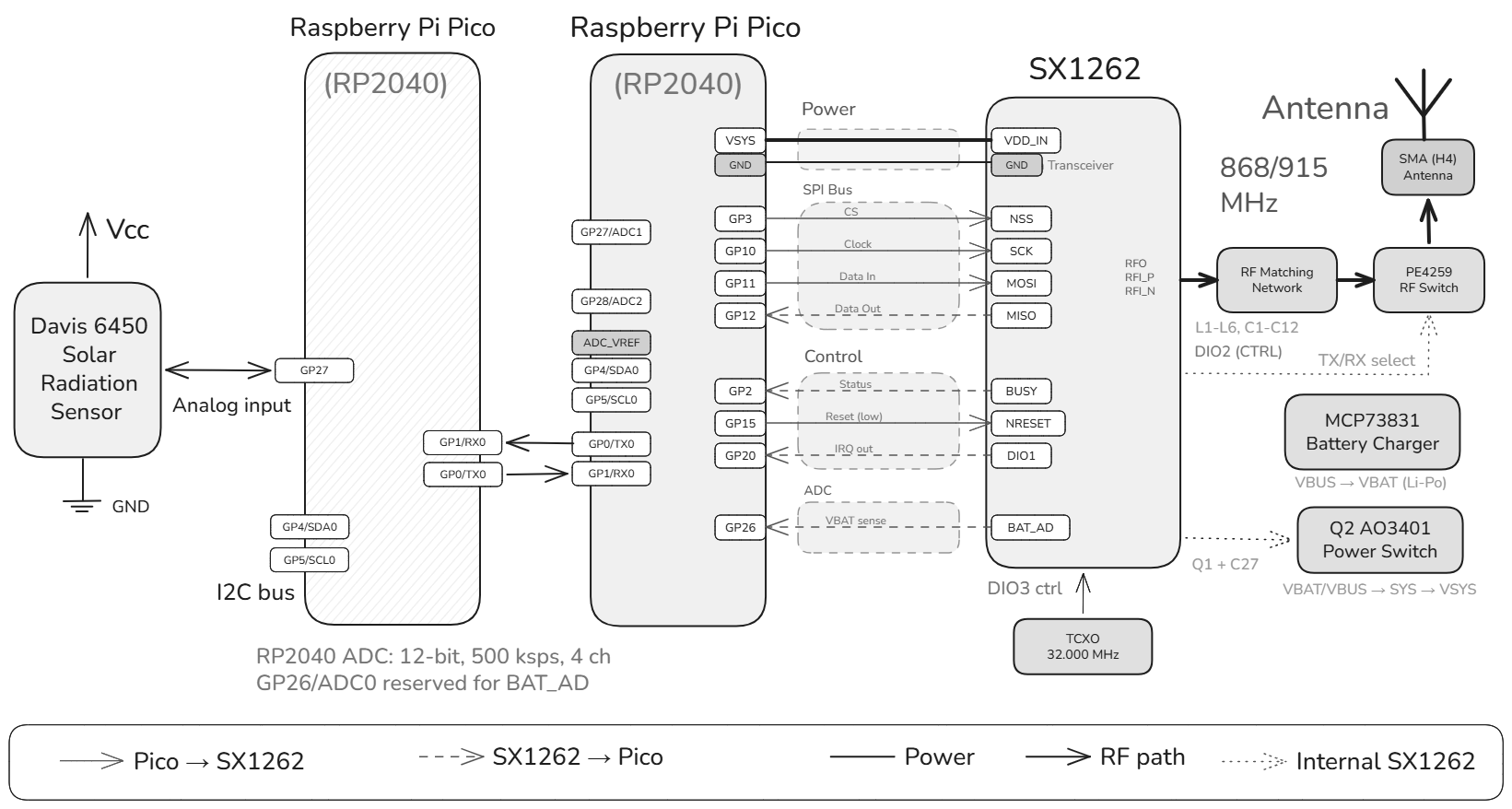}
    \caption{System block diagram of the low-power Meshtastic node (Node 1) based on the RP2040 microcontroller and SX1262 transceiver.}
    \label{fig:sensor_1}
\end{figure}

The perception layer comprises four heterogeneous Meshtastic nodes and one commercial tracking device. Node 1 is a custom-designed, solar-powered unit whose hardware architecture is illustrated in the system block diagram of Fig.~\ref{fig:sensor_1}. This node is built around the Raspberry Pi Pico microcontroller (RP2040, dual-core Cortex-M0+) paired with a Semtech SX1262 LoRa transceiver. It interfaces with a Davis 6450 solar radiation sensor and is intended for continuous, unattended environmental monitoring. 

Its low-power design, based on photovoltaic energy harvesting, removes dependence on grid power and enables permanent outdoor deployment.As shown in the block diagram, the RP2040 interfaces with the SX1262 via a four-wire SPI bus (GP3 for CS, GP10 for SCK, GP11 for MOSI, and GP12 for MISO). Control lines are mapped to specific GPIOs for status handling (BUSY on GP2), reset (NRESET on GP15), and interrupt requests (DIO1 on GP20).The RP2040’s internal 12-bit ADC (500 ksps, four channels) provides analog acquisition on GP27 and GP28 for 0–3.3 V sensors, while GP4 and GP5 (SDA0/SCL0) expose an $I^2C$ bus for digital peripherals. A dedicated UART interface (GP0/GP1) is also available to accommodate serial devices. To optimize power management, GP26/ADC0 is reserved exclusively for battery voltage sensing via the SX1262's BAT\_AD pin, enabling on-node state-of-charge monitoring without external analog conditioning circuitry. In the current deployment, the analog input GP27 is directly connected to the Davis 6450 solar radiation sensor, whose 0–3.3 V output is sampled, converted, and encoded into standard Meshtastic telemetry frames.Power management is handled by an MCP73831 single-cell Li-Po 
charger, which regulates the VBUS supply (from the photovoltaic 
panel) into the battery rail (VBAT). A Q2~AO3401 P-channel MOSFET 
power switch arbitrates between VBAT and VBUS to feed the system 
rail (VSYS), ensuring seamless transitions between solar-powered 
and battery-backed operation. This architecture allows Node~1 to 
sustain unattended outdoor operation without grid power, making 
it suitable for permanent deployment on building rooftops or open 
field infrastructure.

Nodes~2 and~3 are based on the Heltec ESP32 V3 development board, 
which integrates an SX1262 transceiver and supports both LoRa and 
Bluetooth Low Energy (BLE). These nodes function as intermediate 
mesh relays and expose a BLE interface that allows local users to 
interact with the mesh via the Meshtastic mobile application. 
Node~4 is built on the RAK4631 module and operates as the 
network-layer gateway, bridging the LoRa mesh and the local Wi-Fi 
infrastructure. Its role is described in detail in 
Section~\ref{sec:network}.

Mobile asset tracking is provided by a Seeed SenseCAP T1000-E 
device, a compact commercial tracker designed natively for 
Meshtastic firmware. Its core consists of a Nordic nRF52840 
SoC (Bluetooth~5.1) paired with a Semtech LR1110 transceiver, 
which integrates LoRa radio, GNSS scanning, and Wi-Fi scanning 
in a single chip -- enabling multi-technology positioning 
(GPS, GLONASS, Galileo, BeiDou, Wi-Fi, and BLE) without 
external GNSS modules. The device additionally embeds 
temperature, ambient light, and 3-axis accelerometer sensors, 
making it capable of carrying environmental and motion 
telemetry alongside position frames within the same 
Meshtastic packet stream.

The tracker operates from an internal 700~mAh lithium battery 
recharged via a magnetic USB connector, and its IP65-rated 
enclosure (85~$\times$~55~$\times$~6.5~mm, 
$-20$~to~$+60$\textdegree{}C) makes it suitable for 
unattended outdoor deployment across the campus. It 
participates in the mesh as a standard Meshtastic CLIENT node, 
periodically broadcasting \texttt{POSITION\_APP} frames that 
are forwarded hop-by-hop to Node~4 and subsequently published 
to the edge gateway via MQTT for storage and visualization.

\begin{table}[h]
\centering
\caption{Technical specifications of the Seeed Studio SenseCAP T1000-E (Meshtastic Version).}
\label{tab:sensecap_t1000e_specs}
\begin{tabular}{@{}ll@{}}
\toprule
\textbf{Parameter} & \textbf{Technical Specification} \\ \midrule
Microcontroller (MCU) & Nordic nRF52840 (with Bluetooth 5.1) \\
LoRa Transceiver & Semtech LR1110 (LoRa + GNSS/Wi-Fi Scan) \\
Positioning Systems & GNSS (GPS/GLONASS/Galileo/Beidou), Wi-Fi, BLE \\
Internal Sensors & Temperature, Light, 3-axis Accelerometer \\
Battery Capacity & 700 mAh (Rechargeable Lithium) \\
Charging Method & Magnetic USB cable (4.7V - 5.5V DC) \\
Interaction & 1x Function Button, 1x Buzzer, 1x RGB LED \\
Enclosure Rating & IP65 (Waterproof and Dustproof) \\
Dimensions & 85 $\times$ 55 $\times$ 6.5 mm \\
Operating Temperature & -20 \textdegree C to 60 \textdegree C \\ \bottomrule
\end{tabular}
\end{table}

\subsection{Network Layer}
\label{sec:network}

All inter-node communication operates at 915~MHz in the ISM band, 
complying with regional spectrum regulations applicable to the 
Colombian deployment context. The physical-layer configuration 
follows the Meshtastic \textit{LongFast} modem preset, 
corresponding to Spreading Factor SF11 at 125~kHz bandwidth, which 
provides a balance between link budget and data rate adequate 
for campus-scale telemetry.

The routing algorithm is Meshtastic's native \textit{Managed 
Flooding}, a contention-based broadcast mechanism in which each 
node receiving a packet may retransmit it after a randomized 
back-off delay inversely proportional to its received SNR. A 
packet ID cache suppresses duplicate retransmissions, bounding 
network overhead without requiring explicit routing table 
maintenance. This design trades optimal path selection for 
implementation simplicity and resilience to node mobility, which is 
appropriate for a heterogeneous campus deployment where topology 
may change as nodes are relocated.

Node~4 (RAK4631) plays a structurally critical role: it operates 
simultaneously as a LoRa mesh participant and as a Wi-Fi client, 
forwarding decoded Meshtastic payloads as JSON-encoded MQTT 
messages through the local network access point to the edge server. 
This single-node bridge eliminates the need for a dedicated 
LoRaWAN gateway or network server, preserving the fully 
decentralized nature of the mesh while integrating with standard 
IP-based infrastructure.

\subsection{Edge/Processing Layer}

Edge computation is performed by a Raspberry Pi~4 with a static 
local IP assignment to ensure persistent MQTT broker addressing and 
reliable database writes. The entire data pipeline is deployed as a 
multi-container application using Docker Compose, enabling 
reproducible provisioning and service isolation.

The data flow within the edge server is as follows. MQTT messages 
published by Node~4 arrive at a Mosquitto broker (port~1883). A 
Node-RED flow engine subscribes to the relevant topics, parses the 
JSON payload, applies field extraction and unit conversion, and 
writes the resulting time-stamped records to an InfluxDB time-series 
database (port~8086). A Meshview container provides real-time 
topology visualization by consuming the mesh graph information 
embedded in Meshtastic telemetry frames. All containerized services 
persist their data to a local SSD volume, ensuring continuity across 
reboots.

\begin{table}[h]
\centering
\caption{Containerized services in the edge layer.}
\label{tab:edge_services}
\begin{tabular}{@{}llc@{}}
\toprule
\textbf{Service} & \textbf{Function} & \textbf{Port} \\ \midrule
Eclipse Mosquitto & MQTT Broker         & 1883 \\
InfluxDB          & Time-series database & 8086 \\
Node-RED          & ETL Flow orchestration & 1880 \\
Grafana           & Visualization        & 3000 \\
Meshview          & Mesh topology viewer & 8000 \\ \bottomrule
\end{tabular}
\end{table}

\subsection{Application Layer}

Processed data is exposed to end users through two Grafana dashboards 
(port~3000): a Solar Radiation Dashboard displaying irradiance 
time series from Node~1, and a Device Tracking Panel rendering the 
position history of the SenseCAP T1000-E. Both dashboards are 
accessible to local clients via browser over the campus LAN, and 
to remote clients over HTTPS, leveraging the Raspberry Pi~4's 
reverse-proxy configuration. This dual-access model supports both 
on-site monitoring by campus operators and remote supervision by 
researchers off-campus.

\section{Test Methodology and Deployment}

\subsection{Physical Deployment Environment}

The system was deployed at the Universidad Militar Nueva Granada 
(UMNG) Cajicá campus, located at 2,559.88~m above sea level in 
the Sabana de Bogotá region, Colombia (4°55'N, 74°01'W). The 
campus covers 75.5~hectares (~755,000~m²), making it one of 
the largest university campuses in Latin America, and comprises 
a heterogeneous built environment: approximately 16,825~m² of 
laboratories and research facilities, 9,881~m² of classrooms 
and auditoriums, and 18,824~m² of welfare and sports areas, 
all embedded within open fields and dense native vegetation 
adjacent to the Río Bogotá riparian protection zone 
(~118,000~m²). This spatial diversity -- reinforced concrete 
multi-story buildings interspersed with open fields and tree 
canopies -- produces a rich propagation environment combining 
LOS and NLOS segments within a single deployment site.

Three propagation zones were identified for experimental 
purposes. The \textit{open campus zone} covers the central 
fields and walkways between buildings, where LOS conditions 
prevail between adjacent nodes. The \textit{built-up zone} 
encompasses the laboratory and academic building cluster, 
where NLOS conditions dominate and diffraction over rooftops 
constitutes the primary propagation mechanism. A third zone, 
denoted the \textit{extended range scenario}, involves a link 
from the campus perimeter (2,559.88~m a.s.l.) westward along 
the Manas-Molino corridor to Mirador La Cumbre -- Altos de 
Paito (2,628.03~m a.s.l.), located 2.47~km from Node~4. The 
68.15~m elevation advantage of this site over the campus 
gateway provided quasi-LOS propagation conditions over the 
final segment of the route, making it a candidate location 
for a future ROUTER-role relay node in a wider Smart Campus 
mesh deployment.

\subsection{Node Placement and Role Assignment}


The deployment of the network nodes followed a coverage-first strategy, informed by preliminary Received Signal Strength Indicator (RSSI) survey measurements to ensure link reliability across the campus. Table~\ref{tab:nodes} summarizes the hardware specifications, assigned Meshtastic roles, and specific physical locations for each network element.


\begin{table}[htbp]
\centering
\caption{Node Configuration and Deployment Summary}
\label{tab:nodes}
\begin{tabularx}{\textwidth}{@{}cllX@{}} 
\toprule
\textbf{Node} & \textbf{Hardware} & \textbf{Role} & \textbf{Location} \\ \midrule
1 & Pi Pico + SX1262 & CLIENT & Rooftop -- Faculty of Engineering \\
2 & Heltec ESP32 V3 & ROUTER & Main Roundabout -- West Campus Access \\
3 & Heltec ESP32 V3 & ROUTER & Central Academic Wing (Camacho Leiva) \\
4 & RAK4631 & GATEWAY & Laboratory Complex -- South Access  \\
T & SenseCAP T1000-E & TRACKER & Mobile -- User-carried device \\ \bottomrule
\end{tabularx}
\end{table}

Node~1 was strategically mounted on an exposed rooftop at the Faculty of Engineering to maximize solar irradiance on its photovoltaic panel and maintain a clear Line-of-Sight (LoS), thereby minimizing signal attenuation. Node~4 serves as the primary Wi-Fi Gateway and was co-located with a campus access point in the laboratory area near the South Access. This placement was chosen to minimize the network hop distance between the LoRa-to-IP bridge and the centralized edge server, ensuring low-latency data backhaul.

\subsection{Radio Frequency Configuration}

All nodes were configured with identical physical-layer parameters to ensure interoperability within the Meshtastic mesh. Table~\ref{tab:rfconfig} lists the key radio settings applied through the Meshtastic Python CLI and mobile interface prior to deployment.

\begin{table}[htbp]
\centering
\caption{Meshtastic Radio Configuration Parameters}
\label{tab:rfconfig}
\begin{tabular}{@{}ll@{}}
\toprule
\textbf{Parameter} & \textbf{Value} \\ \midrule
Operating Frequency & 915 MHz (US Region) \\
Modem Preset & \textit{LongFast} \\
Spreading Factor (SF) & 11 \\
Bandwidth (BW) & 125 kHz \\
Coding Rate (CR) & 4/5 \\
TX Power & 22 dBm \\
Duty Cycle Limit & Disabled (100\%) \\
Hop Limit & 3 \\ \bottomrule
\end{tabular}
\end{table}

The \textit{LongFast} preset was selected as a compromise between the extended coverage required to bridge the built-up campus zones and the data throughput needed for periodic GPS position frames (transmitted every 60~s by the SenseCAP T1000-E) and solar irradiance samples (logged every 300~s by Node~1). SF12 (\textit{LongSlow}) was evaluated in preliminary tests but discarded due to excessive Time-on-Air (ToA) at the required sampling intervals, which could lead to channel congestion.

\subsection{Experimental Scenarios}

Two evaluation scenarios were defined to characterize different 
aspects of system performance.

\textbf{Scenario~A — Continuous Solar Monitoring:} Node~1 
transmitted fixed-period telemetry frames carrying raw irradiance
readings from the Davis~6450 sensor over a 1-week initial
trial period, which served to identify system bugs and optimize
performance prior to permanent, continuous operation.This scenario evaluates end-to-end pipeline 
stability: LoRa link availability, MQTT delivery, Node-RED 
processing, and InfluxDB write integrity.

\textbf{Scenario~B — Extended Range Link:} A single Meshtastic 
node was transported to the summit of Mirador La Cumbre -- 
Altos de Paito (2,628.03~m a.s.l.), approximately 2.47~km from 
Node~4, to establish a long-range link under quasi-LOS conditions. 
RSSI and SNR values reported by the receiving node were logged via 
the Meshtastic serial interface. This scenario assesses the 
feasibility of using adjacent high-ground features as ROUTER-role 
relay nodes to extend mesh coverage beyond the campus perimeter.

\begin{figure}[htbp]
    \centering
    \includegraphics[width=\columnwidth]{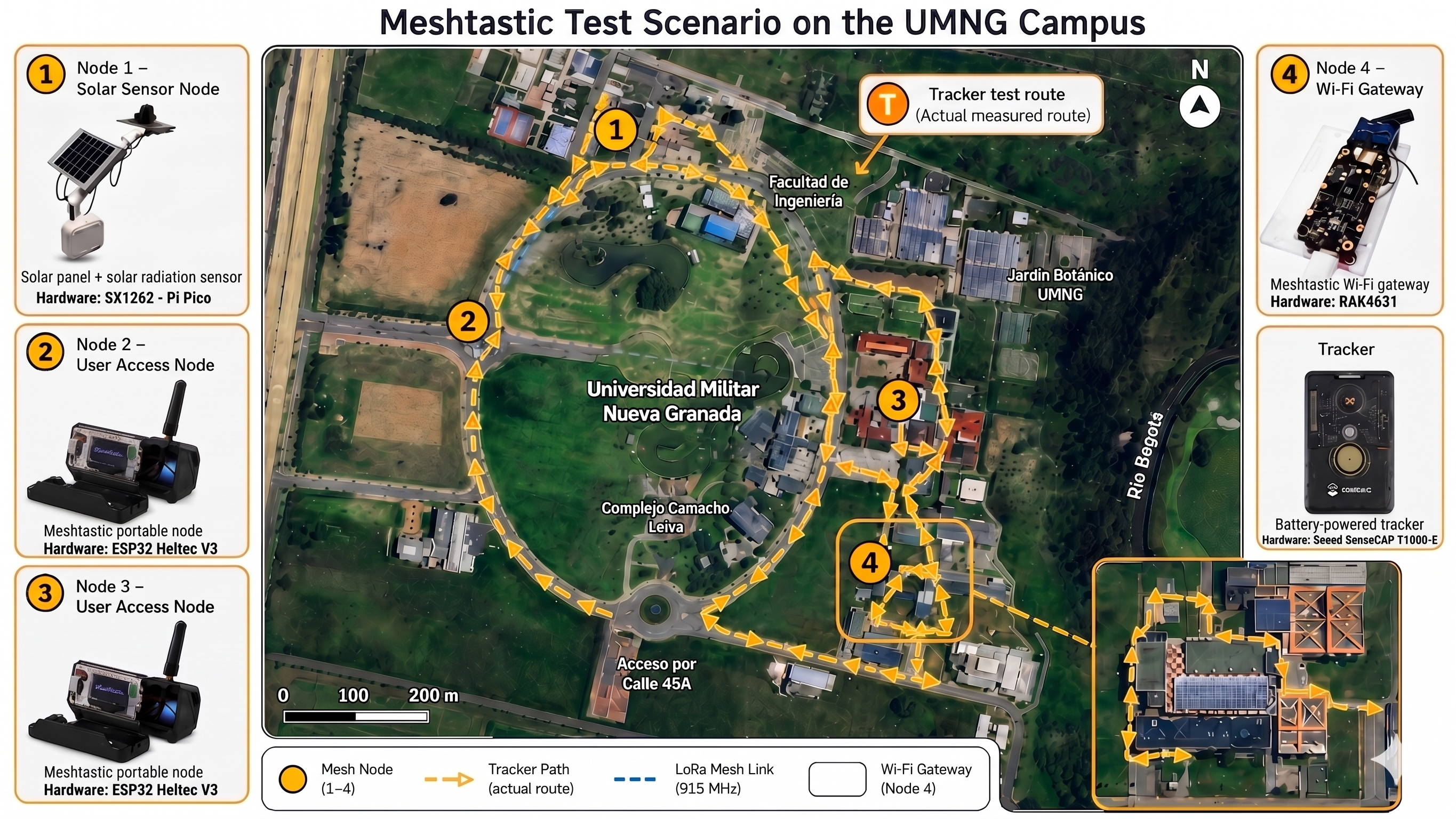}
    \caption{Physical deployment of the Meshtastic mesh on the 
    UMNG Cajicá campus. Orange circles indicate node positions 
    (1--4); the dashed orange path shows the tracker test route; 
    blue dashed lines indicate LoRa mesh links at 915~MHz. The 
    inset detail shows the tracker path within the Node~4 
    laboratory complex.}
    \label{fig:smart_campus}
\end{figure}

Fig.~\ref{fig:smart_campus} shows the physical deployment 
of the four-node mesh on the UMNG Cajicá campus, including 
the tracker test route (Scenario~A) and the position of each 
node relative to the main campus landmarks. The tracker path 
covers the campus perimeter loop at 915~MHz, passing through 
mixed LOS and NLOS segments between the Faculty of Engineering 
(Node~1), the west access roundabout (Node~2), the Camacho 
Leiva academic wing (Node~3), and the laboratory south access 
gateway (Node~4).

\subsection{Extended Range Link: Campus to Mirador La Cumbre}

To assess the viability of using elevated terrain features as 
ROUTER-role relay nodes for expanding mesh coverage, a link test 
was conducted between the fixed gateway (Node~4, RAK4631, 
\textit{Meshtastic Nix}) located at the UMNG Cajicá campus 
(2,559.88~m a.s.l.) and a mobile node (\textit{Meshtastic gaho}, 
Heltec ESP32~V3) carried along a 2.47~km route westward, 
terminating at Mirador La Cumbre -- Altos de Paito 
(2,628.03~m a.s.l.), yielding an elevation advantage of 
68.15~m over the campus gateway.

\begin{figure}[htbp]
    \centering
    \includegraphics[width=\columnwidth]{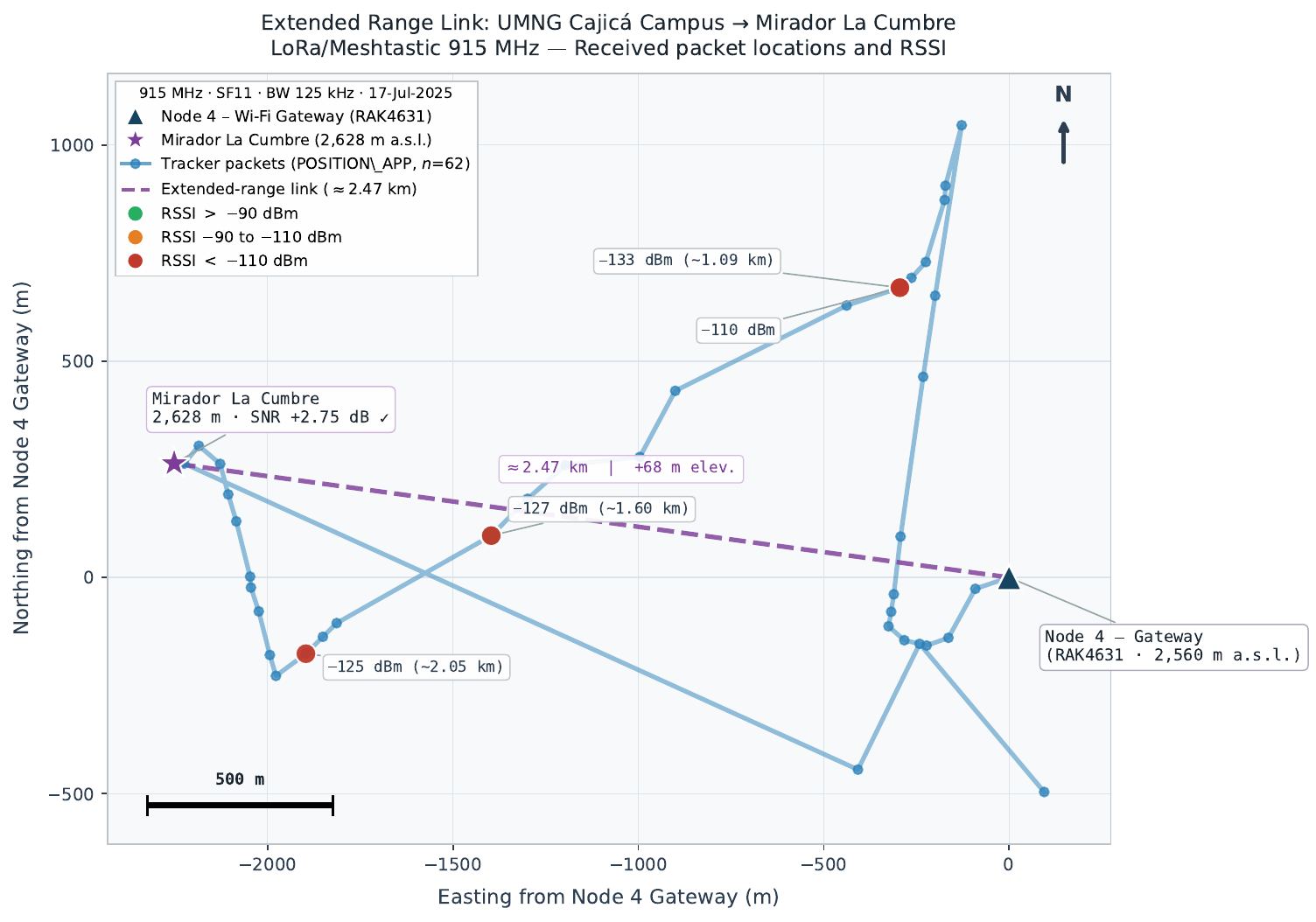}
    \caption{Georeferenced packet reception map for the extended range 
    test (Scenario~B). Axes show displacement in meters from Node~4 
    (gateway). Blue circles indicate POSITION\_APP frames from the 
    SenseCAP T1000-E tracker. Colored markers indicate RSSI-measured 
    packets from \textit{Meshtastic gaho}: green ($>$$-$90~dBm), 
    orange ($-$90 to $-$110~dBm), and red ($<$$-$110~dBm). The 
    dashed purple line shows the 2.47~km extended link to Mirador 
    La Cumbre (2,628~m a.s.l.), where a positive SNR of $+$2.75~dB 
    confirmed decodable reception.}
    \label{fig:ruta_cerro}
\end{figure}

Fig.~\ref{fig:ruta_cerro} shows the georeferenced reception 
map derived from the KML log, comprising 98~packets captured 
across the measurement session. Of these, 62 correspond to 
\texttt{POSITION\_APP} frames from the SenseCAP T1000-E tracker 
(\textit{Meshtastic sebs}), and 36 to control and text frames 
from \textit{gaho} and \textit{Nix}. Table~\ref{tab:rssi_ruta} 
summarizes the RSSI measurements logged by Node~4 at increasing 
distances along the route.

\begin{table}[htbp]
\centering
\caption{RSSI Measurements Along the Campus--Cerro Route}
\label{tab:rssi_ruta}
\begin{tabular}{|c|c|c|}
\hline
\textbf{Approx. Distance (km)} & \textbf{RSSI (dBm)} & 
\textbf{Conditions} \\
\hline
1.09 & $-$110 to $-$133 & Partial NLOS (road corridor) \\
1.60 & $-$124 to $-$127 & NLOS (terrain undulation) \\
2.05 & $-$125             & NLOS (pre-summit approach) \\
2.47 & — (SNR: +2.75~dB) & Quasi-LOS (summit) \\
\hline
\end{tabular}
\end{table}

At the summit of Mirador La Cumbre (2.47~km, +68~m elevation), 
the link remained decodable with a positive SNR of +2.75~dB 
(raw Meshtastic field value: 11, corresponding to 
$11 \times 0.25 = 2.75$~dB), confirming that the SX1262 
receiver margin was sufficient for reliable packet decoding 
despite the extended link distance and absence of direct LOS 
over most of the route. The RSSI variation between $-$110 
and $-$133~dBm observed at 1.09~km reflects the influence 
of terrain undulations and roadside vegetation along the 
Manas-Molino corridor, consistent with mixed LOS/NLOS 
propagation in the Colombian Sabana environment at 2,500+~m 
altitude.

These results support the hypothesis that nodes deployed at 
topographically elevated points -- even with moderate height 
advantage (~68~m) -- can extend Meshtastic mesh coverage 
well beyond the flat campus perimeter. A ROUTER-role node 
at Mirador La Cumbre would provide simultaneous visibility 
to the campus gateway and to areas occluded by campus 
buildings, effectively acting as a mesh backbone node for 
a future Smart Campus extension beyond the UMNG boundary.

\section{Conclusion and Future Work}

\subsection{Conclusion}

This paper presented the design, implementation, and experimental 
evaluation of a LoRa mesh network based on Meshtastic firmware, 
deployed at the UMNG Cajicá campus as a Smart Campus prototype. 
The system integrates heterogeneous sensing nodes, a solar-powered 
custom hardware unit, and a fully containerized edge data pipeline, 
addressing the gap identified in the literature regarding the 
behavior of decentralized Meshtastic deployments under real-world 
heterogeneous conditions.

Two experimental scenarios validated the system across 
complementary operational dimensions. Scenario~A demonstrated 
the end-to-end pipeline integrity of the solar-powered Node~1, 
confirming continuous delivery of solar irradiance telemetry 
from the LoRa PHY layer through Mosquitto, Node-RED, and InfluxDB without interruption. 

The physical deployment across the UMNG Cajicá campus, 
illustrated in Fig.~\ref{fig:smart_campus}, demonstrated 
that the four-node heterogeneous mesh -- combining a 
solar-powered fixed sensor, two ROUTER-role relays, and 
a Wi-Fi gateway -- provides sufficient coverage for the 
campus core loop under mixed LOS/NLOS conditions at 
915~MHz with the \textit{LongFast} preset (SF11, BW~125~kHz).

Scenario~B established that the 
\textit{Managed Flooding} algorithm sustained a decodable 
link at 2.47~km with a positive SNR of $+$2.75~dB at Mirador 
La Cumbre (2,628~m a.s.l.), demonstrating that a modest 
elevation advantage of 68~m is sufficient to establish a 
reliable long-range backbone link over the Sabana de Bogotá 
terrain at 915~MHz with the \textit{LongFast} preset 
(SF11, BW~125~kHz, TX~22~dBm).

Taken together, these results confirm that a Meshtastic-based 
architecture integrated with a Docker Compose microservices 
stack constitutes a viable, low-cost, and reproducible 
platform for Smart Campus IoT deployments, requiring no 
centralized LoRaWAN infrastructure and operating 
autonomously from solar energy.

\subsection{Future Work}

Several directions are identified for extending this work. 
First, the mesh will be expanded with additional 
ROUTER-role nodes at the southern campus perimeter and at 
Mirador La Cumbre, enabling full campus coverage and 
validating the long-range relay hypothesis established in 
Scenario~B. Second, the sensing layer will be extended with new node types targeting rain-induced flood monitoring and Bogotá River level sensing, increasing the heterogeneity of the deployment and its relevance to the broader Smart Campus vision. Third, the 
containerized data pipeline will be augmented with an 
automated decision layer based on InfluxDB task scripts 
and Node-RED alert flows, enabling closed-loop responses 
to sensor thresholds without human intervention. Finally, 
a systematic measurement campaign with controlled 
packet injection will be conducted to obtain statistically 
significant PDR and latency characterizations across all 
campus propagation zones, complementing the qualitative 
coverage assessment reported here.

\bibliographystyle{splncs04}   
\bibliography{references_1}       

\end{document}